\documentclass[preprint,aps,nofootinbib,preprintnumbers,amsmath,amssymb]{revtex4}
\usepackage{graphicx}
\usepackage{dcolumn}
\usepackage{bm}
\usepackage{natbib}
\newcommand{\fft}[2]{{\frac{#1}{#2}}}
\newcommand{\ft}[2]{{\textstyle\frac{#1}{#2}}}
\newcommand{\beq}{\begin{equation}}
\newcommand{\eq}{\end{equation}}
\newcommand{\bea}{\begin{eqnarray}\displaystyle}
\newcommand{\ea}{\end{eqnarray}}

\newcommand{\nn}{\nonumber}

\newcommand{\etas}{\frac{\eta}{s}}
\begin{document}

\preprint{MCTP-09-09}

\title{Higher derivative effects on $\eta/s$ at finite chemical potential}

\author{Sera Cremonini}
\author{Kentaro Hanaki}
\author{James T.~Liu}
\author{Phillip Szepietowski}
\email{{seracre, hanaki, jimliu, pszepiet}@umich.edu}
\affiliation{Michigan Center for Theoretical Physics\\
Randall Laboratory of Physics, The University of Michigan\\
Ann Arbor, MI 48109--1040, USA}

\date{\today}

\begin{abstract}
We examine the effects of higher derivative corrections on $\eta/s$, the
ratio of shear viscosity to entropy density, in the case of a finite
$R$-charge chemical potential.  In particular, we work in the framework of
five-dimensional $\mathcal N =2$ gauged supergravity, and include terms
up to four derivatives, representing the supersymmetric completion of the
Chern-Simons term $A\wedge\mbox{Tr\,}(R\wedge R)$.  The addition of the
four-derivative terms yields a correction which is a $1/N$ effect, and in
general gives rise to a violation of the $\eta/s$ bound.  Furthermore, we
find that, once the bound is violated, turning on the chemical potential
only leads to an even larger violation of the bound.
\end{abstract}

\maketitle

\section{Introduction}

Over the past decade the development of the AdS/CFT
correspondence \cite{Maldacena:1997re,Gubser:1998bc,Witten:1998qj}
has led to a new way of thinking about strongly
coupled gauge theories. Although the original and best studied
example of the AdS/CFT duality connects $\mathcal N=4$
supersymmetric Yang-Mills to type IIB string theory on AdS$_5
\times S^5$, the duality has been extended to a variety of
cases, and can describe confining gauge theories with features
that are qualitatively similar to QCD. In
recent years the AdS/CFT correspondence has proven to be a
valuable tool for better understanding thermal and
hydrodynamic properties of field theories at strong
coupling. In particular, it has been applied to the realm of
heavy ion collisions, with the aim of providing a more realistic
description of the strongly coupled quark-gluon plasma (QGP).

In the context of RHIC physics, a quantity that has played a
special role is the ratio of shear viscosity to entropy
density, $\eta/s$ (see {\it e.g.}~\cite{Gubser:2007zz}
and references therein). Weak coupling calculations in thermal field
theory predict $\eta/s \gg 1$, while elliptic flow measurements
at RHIC seem to indicate a very small ratio $0 \lesssim \eta/s
\lesssim 0.3$, showing that the QGP behaves like a nearly ideal
fluid, and is in the strong coupling regime. Motivated by such
observations, there has been a large effort to apply AdS/CFT
methods to the calculation of various transport coefficients.
The AdS/CFT ``program'' is particularly valuable given that
lattice methods (which work well for equilibrium, or
thermodynamic, quantities) fail for non-equilibrium processes.

Furthermore, developments resulting from the AdS/CFT
correspondence prompted Kovtun, Son and Starinets (KSS) to
postulate a bound \cite{Kovtun:2003wp} for $\eta/s$, according
to which all fluids would obey
\beq \frac{\eta}{s} \geq\frac1{4\pi} \, .
\eq
The bound, which seems to be
satisfied by all substances in nature, was later shown
\cite{Buchel:2003tz} to be saturated in all gauge theories with
a dual supergravity description in the large $N$ and $\lambda=
g_{YM}^2 N$ limit.  Moreover, the universal value
$\eta/s = 1/{4\pi}\sim 0.08$ falls into the experimental range
observed at RHIC.  Finite $\lambda$ corrections to the leading
supergravity result were explored in \cite{Buchel:2004di}, which
considered curvature terms of the form $\sim \alpha'^3 R^4$ in Type IIB
supergravity on $AdS_5 \times S^5$.  The result was that the leading
finite $\lambda$ corrections increase the ratio in the direction consistent
with the bound:
\beq
\frac{\eta}{s} = \frac{1}{4\pi}\Bigl[ 1+ 15 \, \zeta(3)
\lambda^{-3/2} \Bigr] \, .
\eq

However, $\eta/s$ bound violations were subsequently observed in the presence
of curvature squared terms \cite{Kats:2007mq,Brigante:2007nu,Brigante:2008gz,Cai:2009zv}.
In the context of the AdS/CFT correspondence, such terms correspond to
finite $N$ corrections and lead to \cite{Buchel:2008vz}
\beq
\etas =\frac{1}{4\pi}\left(1-\frac{c-a}{a}\right)\, ,
\eq
where $a$ and $c$ are the central charges of the dual CFT.  Thus, violation
will occur provided $c-a>0$.  The central charges are known to be equal in
the large $N$ limit \cite{Henningson:1998gx}, with $a=c=\mathcal O (N^2)$,
but differ for finite $N$.  For the supergravity examples studied so far,
the leading $1/N$ corrections on the CFT side lead to $c-a \geq 0$, implying violation of the bound by finite $N$ effects \cite{Buchel:2008vz}. (It is an
interesting question on its own to ask whether one can have string theory
constructions whose dual description allows for $c-a<0$.)

In this paper we investigate what happens to the $\eta/s$ ratio in the
presence of non-zero chemical potential.  In particular, we focus on the
chemical potential corresponding to turning on a U(1)$_R$ background of
the $\mathcal N=2$ system. To leading order in the
supergravity approximation, the $R$-charge chemical potential does not
affect the calculation of $\eta/s$, as was shown in
\cite{Mas:2006dy,Son:2006em,Maeda:2006by}. However, it is interesting
to examine whether this is still the case once higher derivative corrections
are included. Furthermore, if $\eta/s$ is affected by $R$-charge, it would
be useful to see whether the KSS bound violation gets larger or smaller as
a function of chemical potential%
\footnote{Ideally, one could imagine tuning the chemical potential to match
observations.  However, it should be noted that the $R$-charge chemical
potential we are investigating is not the same as the more physically
relevant chemical potential related to non-zero baryon number density.}.

We work in the framework of $D=5$, $\mathcal N =2$ gauged
supergravity, which is dual to $\mathcal N=1$ super-Yang Mills theory.
In particular, we are interested in supersymmetric
higher derivative terms, which have a highly constrained structure%
\footnote{Four derivative corrections in
the presence of a chemical potential have been partially discussed in
\cite{Ge:2008ni,Cai:2008ph}, where $R^2$ and $F^4$ corrections
were considered, respectively. The supersymmetric Lagrangian, however, has
$RF^2$ and $\nabla F \nabla F$-type terms as well which were not previously
considered.}.
The four-derivative corrections to the leading order supergravity include a
mixed gauge-gravitational Chern Simons term $A \wedge \text{Tr}\,
(R\wedge R)$. The supersymmetric completion of this term was done in
\cite{Hanaki:2006pj}, where an off-shell action was obtained for
$D=5$, $\mathcal N =2$ gauged supergravity at the four-derivative level.
In \cite{Cremonini:2008tw} we derived the corresponding on-shell Lagrangian,
found corrected $R$-charged black hole solutions, and studied their
thermodynamic properties.  We will use many of the results of
\cite{Cremonini:2008tw} to compute the shear viscosity.
Our main result is that turning on $R$-charge not only leads to violation
of the bound, but enhances the effect, pushing $\eta/s$ further below $1/4\pi$.
Furthermore, while the dependence of $\eta$ and $s$ individually on the
$R$-charge is quite complicated, the ratio $\eta/s$ is remarkably simple.

The general picture that emerges from such studies is that if we are interested
in describing properties of the QGP (or other strongly coupled systems), we can
try tuning the parameters available to us (whether $N$, $\lambda$ or the
chemical potential), as long as we remain within the regime of validity
of the supergravity approximation.
Moreover, it is an interesting fundamental question whether violations of the
bound can be related to any constraints on the dual gravitational side or
consistency requirement of the underlying string theory.
For instance, one may be able to relate the violation of the $\eta/s$ bound
to the weak gravity conjecture of \cite{ArkaniHamed:2006dz}, according
to which there should be some states whose $M/Q$ ratio is below the BPS bound.
While this is an interesting avenue to explore%
\footnote{See \cite{Kats:2006xp} for investigating the effect of higher derivatives on
the weak gravity conjecture.},
the solutions that we have considered do not admit a nice extremal BPS black
hole limit (since the extremal solution is the superstar geometry, with a
naked singularity), and therefore do not lend themselves easily to such an
analysis.

The structure of this paper is as follows.  In the following section, we
present the on-shell four-derivative action and write down the non-extremal
$R$-charged background.  Using this as a starting point, we then compute
the shear viscosity in Section~\ref{sec:shear} and conclude with a brief
discussion in Section~\ref{sec:disc}.  Some of the intermediate expressions
are relegated to an appendix.  While this work was being completed
we became aware of \cite{Myers:2009ij}, which overlaps with our results.

\section{$\mathcal N=2$ gauged supergravity and $R$-charged black holes}

Our starting point is five-dimensional $\mathcal N=2$ gauged supergravity.
The physical fields in this theory are the metric $g_{\mu\nu}$, graviphoton
$A_\mu$ and gravitino $\psi_\mu$.  The supersymmetric four-derivative
corrections were obtained in \cite{Hanaki:2006pj} using the superconformal
tensor calculus methods worked out in
\cite{Kugo:2000hn,Bergshoeff:2001hc,Fujita:2001kv,Bergshoeff:2004kh}.
By integrating out the auxiliary fields, the Lagrangian may be put into
the form \cite{Cremonini:2008tw}
\footnote{We follow the conventions of \cite{Hanaki:2006pj} and take
$[\nabla_\mu,\nabla_\nu]v^\sigma= R_{\mu\nu\rho}{}^{\, \sigma} \, v^\rho$ and $R_{ab} = R_{ac\;\;b}^{\;\;\;\,c}$.}
\begin{eqnarray}
\label{OurFinalL}
16\pi G_5{e^{-1}\cal L} &=&-R-\fft14F^{2}
+\fft{1}{12\sqrt{3}}\Bigl(1-4\bar c_2\Bigr)\epsilon^{\mu\nu\rho\lambda\sigma}
A_\mu F_{\nu\rho} F_{\lambda \sigma} + 12g^2\nonumber\\
&&\kern-3em+ \fft{\bar c_2}{g^2}\Bigl[ \frac{1}{16\sqrt{3}} \;
\epsilon_{\mu \nu \rho \lambda \sigma} A^\mu R^{\nu \rho \delta
\gamma} R^{\lambda \sigma}{}_{\delta \gamma} + \frac{1}{8}
\, C_{\mu\nu\rho\sigma}^2
+ \frac{1}{16} C_{\mu \nu \rho \lambda}F^{\mu \nu} F^{\rho \lambda}
- \frac{1}{3} F^{\mu \rho} F_{\rho \nu} R^{\nu}_{\mu} \nn \\
&& - \frac{1}{48} R F^2  + \frac{1}{2} \, F_{\mu
\nu} \nabla^\nu \nabla_\rho F^{\mu \rho} + \frac{1}{4} \,
\nabla^\mu F^{\nu \rho} \nabla_\mu F_{\nu \rho} +
\frac{1}{4} \, \nabla^\mu F^{\nu \rho} \nabla_\nu F_{\rho\mu} \nn \\
&& + \frac{1}{32\sqrt{3}} \; \epsilon_{\mu \nu \rho \lambda \sigma}
F^{\mu \nu} (3 F^{\rho \lambda} \nabla_\delta F^{\sigma \delta} +
4 F^{\rho \delta} \nabla_\delta F^{\lambda \sigma} + 6 F^{\rho}_{\;\;\delta} \nabla^\lambda F^{\sigma \delta})\nn \\
&& +\frac{5}{64} F_{\mu \nu} F^{\nu \rho}F_{\rho
\lambda}F^{\lambda \mu} - \frac{41}{2304}(F^2)^2
\Bigr] \, .
\end{eqnarray}
The four-derivative corrections are determined in terms of a single new
dimensionless parameter $\bar c_2$ (corresponding to $c_2g^2/24$ in the
notation of \cite{Cremonini:2008tw}).  Holographic computation of the Weyl
anomaly \cite{Henningson:1998gx,Blau:1999vz,Nojiri:1999mh,Fukuma:2001uf}
allows $G_5$ and $\bar c_2$ to be expressed in terms of the anomaly
coefficients $a$ and $c$ of the dual $\mathcal N=1$ gauge theory.  This was
worked out in \cite{Buchel:2008vz,Cremonini:2008tw}, with the result
\begin{equation}
g^3G_5=\fft{\pi}{8a},\qquad\bar c_2=\fft{c-a}a.
\label{eq:holoanomaly}
\end{equation}

Nonextremal $R$-charged black hole solutions to the lowest order
$\mathcal{N} = 2$ gauged supergravity were found in \cite{Behrndt:1998jd},
and the corrections linear in $\bar c_2$ were worked out in
\cite{Cremonini:2008tw}.  Using a parameterization convenient for the shear
viscosity calculation, the flat-horizon black holes are given by the metric
\begin{equation}
ds^2=\fft{g^2r_0^2}{u}\Bigl[\fft{f(u)}{H(u)^2}dt^2-H(u)d\vec x\,^2\Bigr]-
\fft{H(u)}{4g^2u^2f(u)}du^2,
\end{equation}
and the gauge field
\begin{equation}
A_t=gr_0\sqrt{\fft{3(1+q)^3}q}\left[1-\fft1{1+qu}-\fft{\bar c_2}2
q(1+q)^3\fft{u^3(1-qu)}{(1+qu)^4}\right].
\label{eq:At}
\end{equation}
The metric functions $f(u)$ and $H(u)$ are given by
\begin{eqnarray}
f&=&(1+qu)^3-(1+q)^3u^2+\bar c_2\left[-\fft83q(1+q)^3u^3
+\fft14(1+q)^6\fft{u^4}{1+qu}\right],\nonumber\\
H&=&1+qu-\fft{\bar c_2}{3}q(1+q)^3\fft{u^3}{(1+qu)^2}.
\label{eq:metfun}
\end{eqnarray}

The above solution is fixed in terms of two parameters, $r_0$ (related to
non-extremality) and dimensionless $q$ (related to the $R$-charge).  At
the two-derivative level, the horizon is located at $u=1$, while the boundary
of AdS$_5$ is at $u=0$.  At linear order in $\bar c_2$, however, the horizon
location gets shifted to
\begin{equation}
u_+=1+\fft{\bar c_2}{12}\fft{(1+q)(3-26q+3q^2)}{2-q}.
\end{equation}
The temperature and entropy density were obtained in \cite{Cremonini:2008tw}
\begin{eqnarray}
T&=&\fft{g^2r_0(2-q)(1+q)^{1/2}}{2\pi}\left[1-\fft{\bar c_2}8
\fft{10-59q-4q^2-3q^3}{(2-q)^2}\right],\nonumber\\
s&=&\fft{(gr_0)^3(1+q)^{3/2}}{4G_5}\left[1+\fft{\bar c_2}8
\fft{21+14q-3q^2}{2-q}\right].
\label{eq:thermo}
\end{eqnarray}
Note that, for $q=0$, we may write the entropy density in terms of the
temperature as
\begin{equation}
s=2\pi^2a\left[1+\fft94\fft{c-a}a\right]T^3,
\end{equation}
where we used the holographic relations (\ref{eq:holoanomaly}).  This
reduces to the familiar $s=\pi^2 N^2 T^3/2$ \cite{Gubser:1996de} for
$\mathcal N=4$ SYM, where $a=c=N^2/4$.

\section{Computation of the shear viscosity}
\label{sec:shear}

We compute the shear viscosity using the Kubo formula, following the
methods developed in \cite{Buchel:2004di,Kats:2007mq}.  In particular,
we introduce a scalar channel perturbation to the metric
\begin{equation}
g_{xy} \rightarrow g_{xy} + h_{xy},
\end{equation}
where, for convenience, we define $h^x{}_y = \phi(t,u,\vec{x})$.  Expanding
the Lagrangian (\ref{OurFinalL}) to second order in the perturbation yields
\begin{eqnarray}
\label{phiaction}
S & = & \frac{1}{16\pi G_5}\int \frac{d^4k}{(2\pi)^4}
\int_{0}^{1}du\Big[ A \phi_k''\phi_{-k} + B\phi_k'\phi_{-k}' + C\phi_k'\phi_{-k} +
D \phi_k\phi_{-k} \nn \\
& & \qquad \qquad \qquad \qquad\qquad + E
\phi_k''\phi_{-k}'' + F \phi_k''\phi_{-k}'\Big],
\end{eqnarray}
where the fourier components of $\phi$ are defined by
\begin{equation}
\phi(t,u,\vec{x}\,)=\int d^3xdt\,\phi_{k}(u)e^{i(\vec{k}\cdot\vec{x}-\omega t)}.
\end{equation}
We note that this parameterization of the action with coefficients
$A,\ldots,F$ was originally used in \cite{Buchel:2004di} to handle the
$R^4$ correction of IIB supergravity.  However, it is general enough to
accommodate the present case.  The coefficients are even functions of the
momentum, and are given explicitly in the appendix.

Varying this action with respect to $\phi$ yields a fourth order
differential equation.  However, since the higher derivative terms are
multiplied by $\bar c_2$, we may reduce the order of the equation by
working perturbatively in $\bar c_2$.  To see this, we first consider
the lowest order equation of motion
\begin{eqnarray}
\label{phieomlo}
\phi'' + \left(\frac{f_0'}{f_0}-\fft1u\right)\phi' +
\frac{\bar\omega^2 H_0^3}{uf_0^2}\phi & = & 0 \, ,
\end{eqnarray}
where we have defined the dimensionless frequency
\begin{equation}
\bar\omega^2=\fft{\omega^2}{4g^4r_0^2}\, .
\end{equation}
The lowest order metric functions
\begin{equation}
f_0=(1+qu)^3-(1+q)^3u^2,\qquad H_0=1+qu,
\end{equation}
are obtained by setting $\bar c_2=0$ in (\ref{eq:metfun}).  Taking
additional derivatives of (\ref{phieomlo}) allows us to eliminate $\phi'''$ and
$\phi''''$ terms in the full equation of motion.  The result is rather
simple:
\begin{equation}
\label{phieom1}
\phi'' + \left(\frac{f'}f-\fft1u-\bar c_2
\frac{(1+q)^3 u}{(1+qu)^3}\right)\phi' + \frac{\bar\omega^2
H^3}{uf^2}\, \phi= 0.
\end{equation}
Notice that the form of this equation is almost identical to that of
(\ref{phieomlo}), the lowest order equation of motion,
modified only by the presence of the corrected metric functions $f$ and $H$ as
well as one new term, which is explicitly ${\cal O}(\bar{c}_2)$.

Since the function $f(u)$ vanishes linearly at the horizon $u_+$, the
point $u=u_+$ is a regular singular point of the equation
of motion (\ref{phieom1}).  This suggests that we write
\begin{equation}
\label{phiansatz} \phi(u) = f(u)^\nu F(u),
\end{equation}
where $F(u)$ is assumed to be regular at the horizon.  The exponent $\nu$
is then obtained by solving the indicial equation.  In the hydrodynamic
limit, the lowest order solution is known \cite{Mas:2006dy,Son:2006em}
and is given by:
\begin{equation}
\label{phi0sol} \phi_0 =
f_0(u)^{\nu_0}\Bigg\{1-\fft{\nu_0}2\Big[\Delta
\ln\frac{(\Xi-\alpha_1-1+2\alpha_3u)(\Xi+\alpha_1+1)}{(\Xi+\alpha_1+1-2\alpha_3u)(\Xi-\alpha_1-1)}
+ 3\ln\big(1+(\alpha_1+1)u-\alpha_3u^2\big) \Big]\Bigg\},
\end{equation}
where
\begin{equation}
\alpha_1 \equiv 3q,\qquad \alpha_2 \equiv
3q^2, \qquad\alpha_3 \equiv q^3,\qquad
\Xi \equiv (1+q)(1+4q)^{1/2},\qquad\Delta \equiv -3\frac{q+1}{\Xi}.
\end{equation}
The exponent $\nu_0$ is given by
\begin{equation}
\nu_0=-\fft{i\bar\omega}{(2-q)(1+q)^{1/2}},
\end{equation}
and may be re-expressed as $\nu_0=-i\omega/4\pi T_0$, where $T_0$ is the
lowest order temperature given in (\ref{eq:thermo}).  Note that we have
chosen incoming wave boundary conditions at the horizon as appropriate to
the shear viscosity calculation.

Adding higher derivative terms will have two effects on this
solution, one being a correction to the function $F(u)$ and
the other a modification of the exponent $\nu$ defined above.  For the
exponent, solving the indicial equation gives
\begin{equation} \label{nu}
\nu = -\frac{i\bar\omega}{(2-q)(1+q)^{1/2}}\Big(1 +
\fft{\bar c_2}8\frac{10-59q-4q^2-3q^3}{(q-2)^2}\Big) =
-\frac{i\omega}{4\pi T},
\end{equation}
where the relation to the temperature (\ref{eq:thermo}) is valid to linear
order in $\bar c_2$.  We may now substitute $\phi(u)=f(u)^\nu F(u)$ into the
equation of motion (\ref{phieom1}) and linearize in $\bar c_2$ to obtain
an equation for $F(u)$.  While this is difficult to solve exactly,
since we only need a solution in the hydrodynamic regime, it is sufficient
to work to first order in $\omega$ (or equivalently $\nu$).
The solution for $F(u)$ is quite complicated and can be found
in the appendix.

Given this solution, it remains to evaluate the on-shell value of the action.
As explained in \cite{Buchel:2004di}, the bulk action (\ref{phiaction})
must be paired with an appropriate generalization of the Gibbons-Hawking term.
In general, the fourth order equation of motion yields a boundary value
problem for the two-point function where additional data must be
specified ({\it e.g.} fields and their first derivatives at the endpoints).
However, when working perturbatively in $\bar c_2$, the equation of motion
reduces to a second order one, given by (\ref{phieom1}).  This allows us to
use a generalized Gibbons-Hawking term of the form
\begin{equation}
\mathcal{K} = - A \phi_k\phi_{-k}' - \frac{F}{2} \phi_k'\phi_{-k}' +
E(p_1\phi_k' + 2p_0\phi_k)\phi_{-k}',
\end{equation}
where
\begin{equation}
p_1=\fft{f_0'}{f_0}-\fft1u,\qquad p_2=\fft{\bar\omega^2H_0^3}{uf_0^2}
\end{equation}
are the coefficients in the lowest order equation of motion (\ref{phieomlo}).

Evaluating the on-shell action then amounts to evaluating a boundary term
\begin{equation}
S=\int\frac{d^4k}{(2\pi)^4}\mathcal{F}_k \Big|_0^1\,,
\end{equation}
where
\begin{eqnarray}
\mathcal{F}_k & = & \frac{1}{16\pi G_5}\biggl[\Bigl(B - A-\fft{F'}2\Bigr)
\phi_k'\phi_{-k} + \frac{1}{2}(C-A')\phi_k\phi_{-k} - E' \phi_k''\phi_{-k}
\nonumber\\
&&\kern4em + E \phi_k''\phi_{-k}'- E\phi_k'''\phi_{-k}
-E\Big(\frac{f_0'}{f_0}-\fft1u\Big)\phi_k'\phi_{-k}' +
2E\frac{\bar\omega^2 H_0^3}{uf_0^2}\phi_k'\phi_{-k} \biggr].
\end{eqnarray}
In order to compute the shear viscosity we need only the limit of
the above action as $u$ approaches the AdS boundary ({\it i.e.}
$u \rightarrow 0$). It turns out that only the
first and third terms contribute.  This yields a value for the shear
viscosity via the Kubo relation
\begin{equation}
\eta=\lim_{\omega\to0}\fft1\omega\lim_{u \rightarrow 0}
(2{\,\rm Im\,}\mathcal{F}_k)
=\frac{(gr_0)^3}{16\pi G_5} \,
(q+1)^{3/2}\Big(1+\fft{\bar c_2}8\frac{5+6q+5q^2}{2-q}\Big).
\end{equation}
Finally, dividing this by the entropy density (\ref{eq:thermo}) gives
a value for the shear viscosity to entropy density ratio of
\begin{equation}
\label{etaovers}
\fft\eta{s} = \frac{1}{4\pi} \Bigl[1 - \bar c_2(1+q) \Bigr]
=\fft1{4\pi}\Bigl[ 1-\fft{c-a}a(1+q) \Bigr],
\end{equation}
where we have rewritten $\bar c_2$ in terms of the anomaly coefficients
$c$ and $a$ using (\ref{eq:holoanomaly}).

\section{Discussion}
\label{sec:disc}

The expression for $\eta/s$, given in (\ref{etaovers}), is surprisingly
simple, given that both $\eta$ and $s$ are individually rather more
complicated functions of the parameter $q$.  This is presumably related to
some form of universality, which holds even in an $R$-charged background%
\footnote{Of course, the simplest result possible would have been to obtain
$\eta/s$ independent of $q$.  But this is clearly not the case here.}.
It is instructive to examine the contribution of the various terms in the
four-derivative action to the result (\ref{etaovers}).  We find that only
four terms in (\ref{OurFinalL}) are important.  Writing
\begin{eqnarray}
16\pi G_5e^{-1}\mathcal L&=&-R-\fft14F^2+\cdots
+\fft{\bar c_2}{g^2}\Bigl[\alpha_1C_{\mu\nu\rho\sigma}^2
+\alpha_2C_{\mu\nu\rho\sigma}F^{\mu\nu}F^{\rho\sigma}\nonumber\\
&&\kern8em+\alpha_3\nabla^\mu F^{\nu\rho}\nabla_\mu F_{\nu\rho}
+\alpha_4\nabla^\mu F^{\nu\rho}\nabla_\nu F_{\rho\mu}+\cdots\Bigr],
\label{eq:a1a4}
\end{eqnarray}
we may arrive at the result
\begin{equation}
\label{etaoversarb}
\fft\eta{s} = \frac{1}{4\pi}
\Bigl[1 -4\bar c_2\bigl(2\alpha_1-
q (\alpha_1+6\alpha_2-6\alpha_3+3\alpha_4)\bigr)\Bigr] \, .
\end{equation}
Note that setting $\alpha_i$ to their actual values in (\ref{OurFinalL})
reproduces (\ref{etaovers}).

The shear viscosity to entropy density ratio was independently derived
in \cite{Myers:2009ij}, where it was found to depend only on terms
explicitly involving the Riemann tensor [{\it i.e.} the $\alpha_1$ and
$\alpha_2$ terms in (\ref{eq:a1a4})].  This appears to differ from the
result found above.  However, by the use of Bianchi identities
and integration by parts we can cast the gradient terms into the form
\begin{eqnarray}
\alpha_3\nabla^\mu F^{\nu\rho}\nabla_\mu F_{\nu\rho} +
\alpha_4\nabla^\mu F^{\nu\rho}\nabla_\nu F_{\rho\mu} &= & \nonumber \\
&&\kern-10em
(2\alpha_3-\alpha_4) \Big[-F_{\mu\nu}\nabla^\nu\nabla_\rho F^{\mu\rho}
+F^{\mu\rho}F_{\rho\nu}R_\mu^\nu
-\ft12R_{\mu\nu\rho\sigma}F^{\mu\nu}F^{\rho\sigma}\Big].
\end{eqnarray}
The first two terms do not contribute to the $\eta/s$ ratio, while the last
term will add to the original $\alpha_2$ term to give an effective
$\tilde\alpha_2 = \alpha_2 -\alpha_3 +\alpha_4/2$, so that (\ref{etaoversarb})
may be rewritten as
\begin{equation}
\fft\eta{s} = \frac{1}{4\pi}
\Bigl[1 -4\bar c_2\bigl(2\alpha_1-q (\alpha_1+6\tilde\alpha_2)\bigr)\Bigr] \, .
\end{equation}
This agrees with the result of \cite{Myers:2009ij}
provided the difference in signature conventions is taken into account.

Finally, we return to the $\mathcal N=1$ SYM shear viscosity result of
(\ref{etaovers}).  In order to express this in terms of physical quantities,
we wish to relate the parameter $q$ to the $R$-charge chemical
potential and temperature.  Since $q$ only enters into (\ref{etaovers}) at
the next-leading order, we can use the leading order expressions in pinning
down $q$.  The chemical potential for $R$-charge $\Phi$ is identified as
the difference of $A_t$ between horizon and boundary
\cite{Chamblin:1999tk, Cvetic:1999ne}. At lowest order, (\ref{eq:At}) yields
\begin{equation}
\Phi = g r_0 \sqrt{3q(1+q)}.
\end{equation}
Comparing this to the temperature
\begin{equation}
T_0=\fft{g^2r_0}{2\pi}(2-q)(1+q)^{1/2},
\end{equation}
allows us to write
\begin{equation}
q=\fft3{2\bar\Phi^2}\left(1+\fft43\bar\Phi^2-\sqrt{1+\fft83\bar\Phi^2}\right),
\label{eq:qsol}
\end{equation}
where $\bar\Phi=g\Phi/2\pi T$ is the dimensionless chemical potential.  Note
that $q$ is an increasing function with respect to $\bar\Phi$, with $q=0$ when
$\bar\Phi=0$.  The possible value of $q$ ranges as
\begin{equation}
0 \leq q \leq 2.
\label{eq:qrange}
\end{equation}
Substituting (\ref{eq:qsol}) into (\ref{etaovers}) then gives
\begin{equation}
\frac{\eta}{s} = \frac{1}{4\pi}\left[1 -
\frac{c-a}{a}\left(1+\fft3{2\bar\Phi^2}\left(1+\fft43\bar\Phi^2-\sqrt{1+\fft83\bar\Phi^2}\right) \right) \right].
\end{equation}
Since $q$ is non-negative, this demonstrates that turning on an $R$-charge
chemical potential only increases violation of the $\eta/s$ bound, provided
$c-a>0$.  Taking the range (\ref{eq:qrange}) into account, we see that
adjusting the $R$-charge yields a range of values
\begin{equation}
\fft1{4\pi}\left(1-3\fft{c-a}a\right)\le\fft\eta{s}\le
\fft1{4\pi}\left(1-\fft{c-a}a\right),
\end{equation}
where we have again assumed $c-a>0$.

In conclusion, we have explored the effect of a background $R$-charge on
the shear viscosity to entropy density ratio $\eta/s$.  While the leading
order ratio $\eta/s=1/4\pi$ is universal, $R$-charge corrections do turn
up at the $1/N$ order.  For known theories with a holographic dual,
where $c-a>0$, the conjectured $1/4\pi$ bound is generally violated for
arbitrary chemical potential.  We caution, however, that this is a
parametrically small violation appearing at $\mathcal O(1/N)$ in the
large $N$ limit.  In principle, it would be desirable to obtain a more
robust result.  However, this is hindered by difficulties in obtaining
exact solutions to the full equations of motion ({\it i.e.} beyond the
linearized limit).  While this can be done in certain cases such as
Gauss-Bonnet gravity, the natural supersymmetric organization of the
higher derivative Lagrangian (\ref{OurFinalL}) is not of this form.
It would be interesting to see if a modified universality relation for
$\eta/s$ can be obtained for arbitrary forms of the higher derivative gravity
theory.

\section{Acknowledgments}
We would like to thank A. Buchel, R. Myers and A. Sinha for valuable discussions.
We are particularly grateful to A. Sinha for bringing their
work \cite{Myers:2009ij} to our attention at the concluding stages of this
project. SC would like to thank McGill University for their hospitality
during the final phase of this work.
This work is supported in part by the US Department of Energy under grant
DE-FG02-95ER40899.

\section{Appendix}

The quadratic action for the scalar channel perturbation $\phi$ is
given in (\ref{phiaction}) in terms of six coefficients $A,\ldots,F$.
Here we present their explicit forms:
\begin{eqnarray}
A(u)& = & \frac{4}{u}f_0 +
\bar c_2\bigg[-\fft{\omega^2}{g^2}\frac{H_0^2}{3}+
\frac{2uf_0(1+q)^3(5qu-1)}{H_0^3} -
\frac{32g^2qu^2(1+q)^3}{3} +\frac{g^2u^3(1+q)^6}{H_0} \bigg], \nonumber \\
B(u) &=& \frac{3f_0}{u} + \bar c_2\bigg[-\fft{\omega^2}{g^2}H_0^2 +
\frac{g^2(4qu+1)^2H_0^3}{3u} -
\frac{g^2u(1+q)^3(56q^2u^2+7qu+11)}{6} \nonumber \\
&& \kern3.5em+ \frac{g^2u^3(1+q)^6(26q^2u^2-17qu+17)}{6H_0^3} -
8g^2(1+q)^3qu^2 + \frac{3g^2u^3(1+q)^6}{4H_0}\bigg], \nonumber \\
C(u) &=& \frac{2g^2(4qu-3)H_0^2}{u^2} -
\frac{2g^2(1+q)^3(2qu+1)}{H_0}  \nonumber \\ &&+
\bar c_2\bigg[-\frac{\omega^2}{6uf_0}\Big((4qu+1)H_0^4
-(1+q)^3(-11qu^3+13u^2)H_0 \Big)  \nonumber \\
&& \kern1.3em - \frac{g^2(1+q)^3(4q^2u^2+45qu+3)}{3H_0} +
\frac{g^2u^2(1+q)^6(4q^3u^3-7q^2u^2-32qu+15)}{2H_0^4} \bigg],\nonumber \\
D(u) &=& \frac{2g^2H_0^3-g^2qu^3(1+q)^3}{u^3H_0^2} +
\omega^2\frac{H_0^3}{4u^2f_0} \nonumber \\ &&\! +
\bar c_2\bigg[\fft{\omega^4}{g^2}\frac{H_0^5}{12uf_0^2} +
\frac{\omega^2g^2(1+q)^3}{48f_0^2}\Big(2(31qu-9)H_0^3
-3u^2(1+q)^3(5q^2u^2-4qu+11) \Big) \nonumber \\ && \qquad -
\frac{19g^2q(1+q)^3}{3H_0^2} -
\frac{3g^2u(1+q)^6(6q^2u^2-17qu+1)}{2H_0^5} \bigg], \nonumber \\
E(u) &=& \bar c_2\frac{4uf_0^2}{3g^2H_0}, \nonumber \\
F(u) &=& \bar c_2\, f_0\, \frac{2(2(4qu+1)H_0^3 -
u^2(1+q)^3(7qu+4))}{3H_0^2}.
\end{eqnarray}

Here we also present the  $\mathcal{O}(\bar c_2)$ solution for $\phi$.
Writing $\phi(u) = f(u)^\nu F(u)$, we may expand $F(u)$ to first order
in both $\bar c_2$ and $\omega$
\begin{equation}
F(u) = F_0(u,\omega) + \bar{c_2}(F_{10}(u) + \omega F_{11}(u)).
\end{equation}
Since $F(u)$ satisfies a second order equation (after linearizing in
$\bar c_2$ and using the lowest order equation of motion), it is consistent
to choose the boundary conditions such that $F(u)$ is normalized at
the boundary ($F(0)=1$) and is regular at the horizon.

The function $F_0(u,\omega)$ is given by the expression in the curly
brackets in (\ref{phi0sol}), while the remaining functions are
\begin{eqnarray}
F_{10}(u) &=& 0, \nonumber \\
F_{11}(u) &=&
\frac{(1+q)^{3/2}(11q^5+4q^4+179q^3-10q^2-8q-16)}{32q^2(1+q)^2(q-2)^3}\Big[i
\ln(q^3u^2-3qu-u-1) + \pi\Big] \nonumber \\ & & +
\frac{i(q+1)^{3/2}(60q^6+99q^5+648q^4-69q^3-154q^2-104q-16)}{16(4q+1)^{3/2}(q+1)^2(q-2)^3}
\times \nonumber \\ && \kern4em \left[
\tanh^{-1}\frac{-(1+3q)}{(4q+1)^{1/2}(q+1)} -
\tanh^{-1}\frac{2q^3u-(1+3q)}{(4q+1)^{1/2}(q+1)} \right] \nonumber \\
&& - \frac{i\ln(1+qu)(1+q)^{3/2}}{8q^2} -
\frac{i(q+1)^{3/2}(-4q^5+21q^4+143q^3-21q^2-39q-6)}{8q^4(4q+1)(q-2)^2}
\nonumber \\
&&
-\frac{i(q+1)^{3/2}(4q^7-27q^6+64q^5+511q^4+137q^3-128q^2-57q-6)qu^2}{8(1+qu)q^4(q^3u^2-3qu-u-1)(4q+1)(q-2)^2}
\nonumber \\
&&  +
\frac{i(-12q^6+102q^5+605q^4+63q^3-177q^2-63q-6)u}{8(1+qu)q^4(q^3u^2-3qu-u-1)(4q+1)(q-2)^2}
\nonumber \\
&&  - \frac{i(4q^5+21q^4+143q^3-21q^2-39q-6)
}{8(1+qu)q^4(q^3u^2-3qu-u-1)(4q+1)(q-2)^2}.
\end{eqnarray}


\end{document}